\documentclass[twocolumn,showpacs,preprintnumbers,amsmath,amssymb]{revtex4}

\usepackage{graphicx}
\usepackage{dcolumn}
\usepackage{bm}

\def\ii{{\mathrm{i}}}

\def\ee{{\mathrm{e}}}
\def\dd{{\mathrm{d}}}

\def\bracket#1{\langle #1 \rangle}

\def\diver{\mathop{\mathrm{div}}}
\def\rot{\mathop{\mathrm{rot}}}

\def\sub#1{_{\mathrm{#1}}}
\def\up#1{^{\mathrm{#1}}}
\def\Vec#1{\mbox{\boldmath $#1$}}

\begin{document}

\preprint{APS/123-QED}

\title{Kolmogorov spectrum of superfluid turbulence: numerical analysis of the Gross-Pitaevskii equation with the small
 scale dissipation}

\author{Michikazu Kobayashi}
\author{Makoto Tsubota}%
\affiliation{Department of Physics, Osaka City University, Sumiyoshi-Ku, Osaka 558-8585, Japan}%


\date{\today}

\begin{abstract}
The energy spectrum of superfluid turbulence is studied numerically by solving the Gross-Pitaevskii equation.
We introduce the dissipation term which works only in the scale smaller than the healing length, to remove short
 wavelength excitations which may hinder the cascade process of quantized vortices in the inertial range.
The obtained energy spectrum is consistent with the Kolmogorov law.
\end{abstract}

\maketitle

The physics of quantized vortices in liquid $^4$He is one of the most important topics in low temperature
 physics \cite{Donnelly}.
Liquid $^4$He enters the superfluid state at 2.17 K.
Below this temperature, the hydrodynamics is usually described using the two-fluid model in which the system consists
 of inviscid superfluid and viscous normal fluid.
Early experimental works on the subject focused on thermal counterflow in which the normal fluid flowed in the opposite
 direction to the superfluid flow.
This flow is driven by the injected heat current, and it was found that the superflow becomes dissipative when the
 relative velocity between two fluids exceeds a critical value \cite{Donnelly}.
Feynman proposed that this is a superfluid turbulent state consisting of a tangle of quantized vortices \cite{Feynman}.
Vinen confirmed Feynman's picture experimentally by showing that the dissipation comes from the mutual friction between
 vortices and the normal flow \cite{Vinen1}.
After that, many studies on superfluid turbulence (ST) have been devoted to thermal counterflow \cite{Tough}.
However, as thermal counterflow has no analogy with conventional fluid dynamics, we have not understood the relation
 between ST and classical turbulence (CT).
With such a background, experiments of ST that do not include thermal counterflow have recently appeared.
Such studies represent a new stage in studies of ST.
Maurer {\it et al}. studied a ST that was driven by two counter-rotating disks \cite{Maurer} in superfluid $^4$He at
 temperatures above 1.4 K, and they obtained the Kolmogorov law.
Stalp {\it et al}. observed the decay of grid turbulence \cite{Stalp} in superfluid $^4$He at temperatures above 1 K,
 and their results were consistent with the Kolmogorov law.
The Kolmogorov law is one of the most important statistical laws \cite{Frisch} of fully developed CT, so these
 experiments show a similarity between ST and CT.
This can be understood using the idea that the superfluid and the normal fluid are likely to be coupled together by the
 mutual friction between them and thus to behave like a conventional fluid \cite{Vinen2}.
Since the normal fluid is negligible at very low temperatures, an important question arises: even without the normal
 fluid, is ST still similar to CT or not?

To address this question, we consider the statistical law of CT \cite{Frisch}.
The steady state for fully developed turbulence of an incompressible classical fluid follows the Kolmogorov law for the
 energy spectrum.
The energy is injected into the fluid at some large scales in the energy-containing range.
This energy is transferred in the inertial range from large to small scales without been dissipated.
The inertial range is believed to be sustained by the self-similar Richardson cascade in which large eddies are
 broken up to smaller ones, having the Kolmogorov law
\begin{equation}
E(k)=C\epsilon^{2/3}k^{-5/3}. \label{eq-Kolmo}
\end{equation}
Here the energy spectrum $E(k)$ is defined as $E=\int\dd k\: E(k)$, where $E$ is the kinetic energy per unit mass and
 $k$ is the wave number from the Fourier transformation of the velocity field.
The energy transferred to smaller scales in the energy-dissipative range is dissipated by the viscosity with the
 dissipation rate, which is identical with the energy flux $\epsilon$ of Eq. (\ref{eq-Kolmo}) in the inertial range.
The Kolmogorov constant $C$ is a dimensionless parameter of order unity.

In CT, the Richardson cascade is not completely understood, because it is impossible to definitely identify each eddy.
In contrast, quantized vortices in superfluid are definite and stable topological defects.
A Bose-Einstein condensed system yields a macroscopic wave function
 $\Phi(\Vec{x},t)=\sqrt{\rho(\Vec{x},t)}\ee^{\ii\phi(\Vec{x},t)}$, whose dynamics is governed by the Gross-Pitaevskii
 (GP) equation \cite{Gross,Pitaevskii}
\begin{equation}
\ii\frac{\partial}{\partial t}\Phi(\Vec{x},t)=[-\nabla^2-\mu+g|\Phi(\Vec{x},t)|^2]\Phi(\Vec{x},t),\label{eq-GP}
\end{equation}
where $\mu$ is the chemical potential, $g$ is the coupling constant, $\rho(\Vec{x},t)$ is the condensate density, and
 $\Vec{v}(\Vec{x},t)$ is the superfluid velocity given by $\Vec{v}(\Vec{x},t)=2\nabla\phi(\Vec{x},t)$.
The vorticity $\rot\Vec{v}(\Vec{x},t)$ vanishes everywhere in a single-connected region of the fluid; any rotational
 flow is carried only by a quantized vortex in the core of which $\Phi(\Vec{x},t)$ vanishes so that the circulation is
 quantized by $4\pi$.
The vortex core size is given by the healing length $\xi=1/\sqrt{g\rho}$.
Although quantized vortices at finite temperatures can decay through mutual friction with the normal fluid, at very low
 temperatures the vortices can decay only by two mechanisms; one is the sound emission through vortex
 reconnections \cite{Leadbeater}, and the other is that vortices reduce through the Richardson cascade process and
 eventually change to elementary excitations at the healing length scale.
In any case, dissipation occurs only at scales below the healing length.
Therefore, ST at very low temperatures, consisting of such quantized vortices, can be a prototype to study the
 inertial range, the Kolmogorov law and the Richardson cascade.

There are two kinds of formulation to study the dynamics of quantized vortices; one is the vortex-filament model
 \cite{Schwarz}, and the other the GP model.
By using the vortex-filament model with no normal fluid component, Araki {\it et al}. studied numerically a vortex
 tangle starting from a Taylor-Green flow, thus obtaining the energy spectrum consistent with the Kolmogorov law
 \cite{Araki}.
By eliminating the smallest vortices whose size is comparable to the numerical space resolution, they introduced the
 dissipation into the system.
Nore {\it et al}. used the GP equation to numerically study the energy spectrum of ST \cite{Nore}.
The kinetic energy consists of a compressible part due to sound waves and an incompressible part
 coming from quantized vortices.
Excitations of wavelength less than the healing length are created through vortex reconnections or through the
 disappearance of small vortex loops \cite{Leadbeater,Ogawa}, so that the incompressible kinetic energy transforms into
 compressible kinetic energy while conserving the total energy.
The spectrum of the incompressible kinetic energy is temporarily consistent with the Kolmogorov law.
However, the consistency becomes weak in the late stage when many sound waves created through those processes hinder
 the cascade process \cite{Berloff1,Berloff2,Koplik}.

Our approach here is to introduce a dissipation term that works only on scales smaller than the
 healing length $\xi$.
This dissipation removes not vortices but short wavelength excitations, thus preventing the excitation energy from
 transforming back to vortices.
Compared to the usual GP model, this approach enables us to more clearly study the Kolmogorov law.

To solve the GP equation numerically with high accuracy, we use the Fourier spectral method in space with periodic
 boundary condition in a box with spatial resolution containing $256^3$ grid points.
We solve the Fourier transformed GP equation
\begin{eqnarray}
\ii\frac{\partial}{\partial t}\tilde{\Phi}(\Vec{k},t)&=&[k^2-\mu]\tilde{\Phi}(\Vec{k},t)\nonumber\\
& &+\frac{g}{V^2}\sum_{\Vec{k}_1,\Vec{k}_2}\tilde{\Phi}(\Vec{k}_1,t)\tilde{\Phi}^\ast(\Vec{k}_2,t)\nonumber\\
& &\times\tilde{\Phi}(\Vec{k}-\Vec{k}_1+\Vec{k}_2,t)\label{eq-FtransGP},
\end{eqnarray}
where $V$ is the volume of the system and $\tilde{\Phi}(\Vec{k},t)$ is the spatial Fourier component of
 $\Phi(\Vec{x},t)$ with the wave number $\Vec{k}$, numerically given by a Fast-Fourier-Transformation \cite{Press}.
We consider the case of $g=1$.
For numerical parameters, we used a spatial resolution $\Delta x=0.125$ and $V=32^3$, where the length scale is
 normalized by the healing length $\xi$.
With this choice, $\Delta k=2\pi/32$.
Numerical time evolution was given by the Runge-Kutta-Verner method with the time resolution
 $\Delta t=1\times 10^{-4}$.

\begin{figure}[htb]
\begin{center}
\begin{minipage}{0.49\linewidth}
\begin{center}
\includegraphics[width=0.95\linewidth]{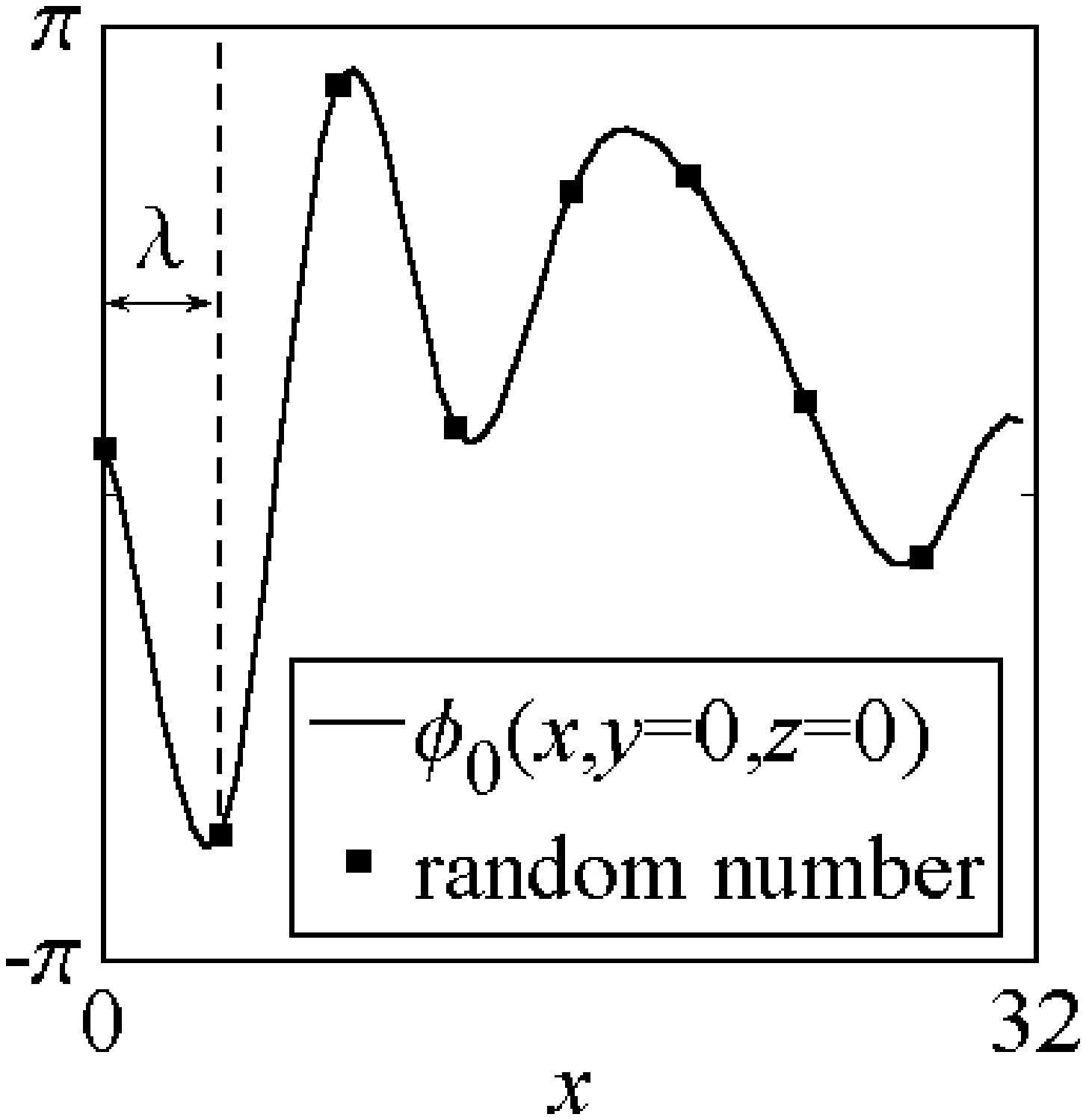}\\
(a)
\end{center}
\end{minipage}
\begin{minipage}{0.49\linewidth}
\begin{center}
\includegraphics[width=0.95\linewidth]{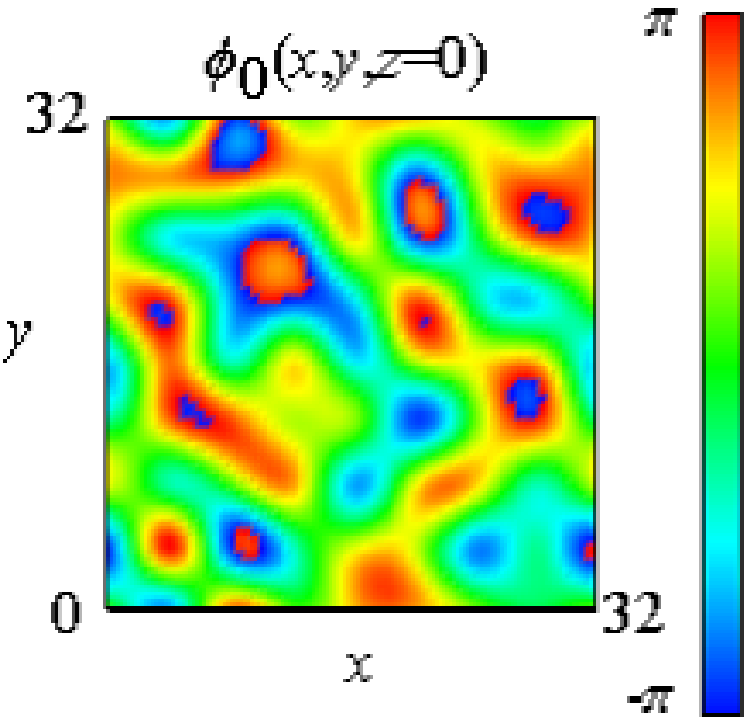}\\
(b)
\end{center}
\end{minipage}
\caption{\label{fig-randpha} How to make the random phase $\phi_0(\Vec{x})$ with $\lambda=4$ (a) and one example
 of an initial random phase in a cross section $\phi_0(x,y,0)$ (b).}
\end{center}
\end{figure}

To obtain a turbulent state, we start from an initial configuration in which the condensate density $\rho_0$ is uniform
 and the phase $\phi_0(\Vec{x})$ has a random spatial distribution.
Here the random phase $\phi_0(\Vec{x})$ is made by placing random numbers between $-\pi$ to $\pi$ at every distance
 $\lambda=4$ and connecting them smoothly (Fig. \ref{fig-randpha}).
The initial velocity $\Vec{v}(\Vec{x},t=0)=2\nabla\phi_0(\Vec{x})$ given by the initial random phase is random, hence
 the initial wave function is dynamically unstable and soon produces homogeneous
 and isotropic turbulence with many quantized vortex loops.

To confirm the accuracy of our simulation, we calculate the total energy $E$, the interaction energy $E\sub{int}$, the
 quantum energy $E\sub{q}$ and the kinetic energy $E\sub{kin}$ \cite{Nore}:
\begin{eqnarray}
 E&=&\frac{\int\dd\Vec{x}\:\Phi^\ast[-\nabla^2+g/2|\Phi|^2]\Phi}{\int\dd\Vec{x}\:\rho}, \nonumber \\
 E\sub{int}&=&\frac{g}{2}\frac{\int\dd\Vec{x}\:|\Phi|^4}{\int\dd\Vec{x}\:\rho}, \nonumber \\
 E\sub{q}&=&\frac{\int\dd\Vec{x}\:[\nabla|\Phi|]^2}{\int\dd\Vec{x}\:\rho}, \nonumber \\
 E\sub{kin}&=&\frac{\int\dd\Vec{x}\:[|\Phi|\nabla\phi]^2}{\int\dd\Vec{x}\:\rho} \label{eq-manyE},
\end{eqnarray}
 and compare them with those given by the dynamics with the different time resolution $\Delta t=2\times 10^{-5}$ or
 different spatial resolutions $512^3$ grids.
For the comparison, we estimate the relative errors $F_{12}(A)=|(\bracket{A}_1-\bracket{A}_2)/\bracket{A}_1|$ and
 $F_{13}(A)=|(\bracket{A}_1-\bracket{A}_3)/\bracket{A}_1|$ at $t=12$ for $A=E$, $E\sub{int}$, $E\sub{q}$ and
 $E\sub{kin}$, where $\bracket{}_1$, $\bracket{}_2$ and $\bracket{}_3$ are respectively the values given by three
 simulations with different resolutions (i) $\Delta t=1\times 10^{-4}$ and $256^3$ grids, (ii)
 $\Delta t=2\times 10^{-5}$ and $256^3$ grids and (iii) $\Delta t=1\times 10^{-4}$ and $512^3$ grids.
Comparison is shown in Table \ref{table-error}; the relative errors are extremely small, which allows us to use the
 resolution (i).
\begin{table}[htb]
\begin{center}
\begin{tabular}{|c|c|c|}\hline
 & $F_{12}(A)$ & $F_{13}(A)$ \\ \hline
$A=E$ & $2.4\times 10^{-12}$ & $6.3\times 10^{-10}$ \\ \hline
$A=E\sub{int}$ & $3.7\times 10^{-12}$ & $8.8\times 10^{-10}$ \\ \hline
$A=E\sub{q}$ & $2.6\times 10^{-12}$ & $6.9\times 10^{-10}$ \\ \hline
$A=E\sub{kin}$ & $5.1\times 10^{-12}$ & $9.4\times 10^{-10}$ \\ \hline
\end{tabular}
\caption{\label{table-error} Dependence of $F_{12}$ and $F_{13}$ on $E$, $E\sub{int}$, $E\sub{q}$ and $E\sub{kin}$.}
\end{center}
\end{table}
Furthermore, the total energy is conserved by the accuracy of $10^{-10}$ in our simulation.

\begin{figure}[htb]
\begin{center}
\begin{minipage}{0.49\linewidth}
\begin{center}
\includegraphics[width=0.95\linewidth]{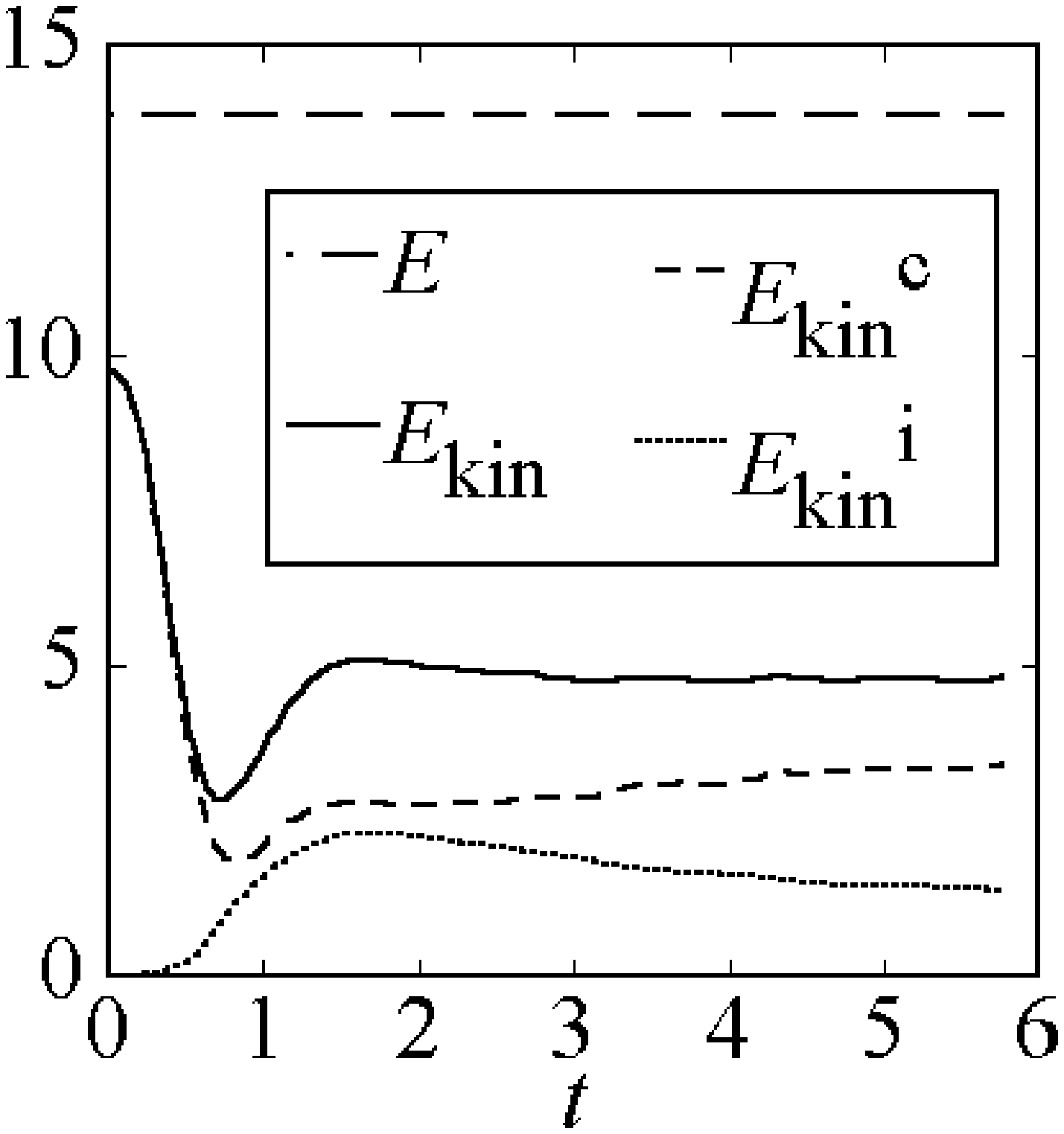}\\
(a)
\end{center}
\end{minipage}
\begin{minipage}{0.49\linewidth}
\begin{center}
\includegraphics[width=0.95\linewidth]{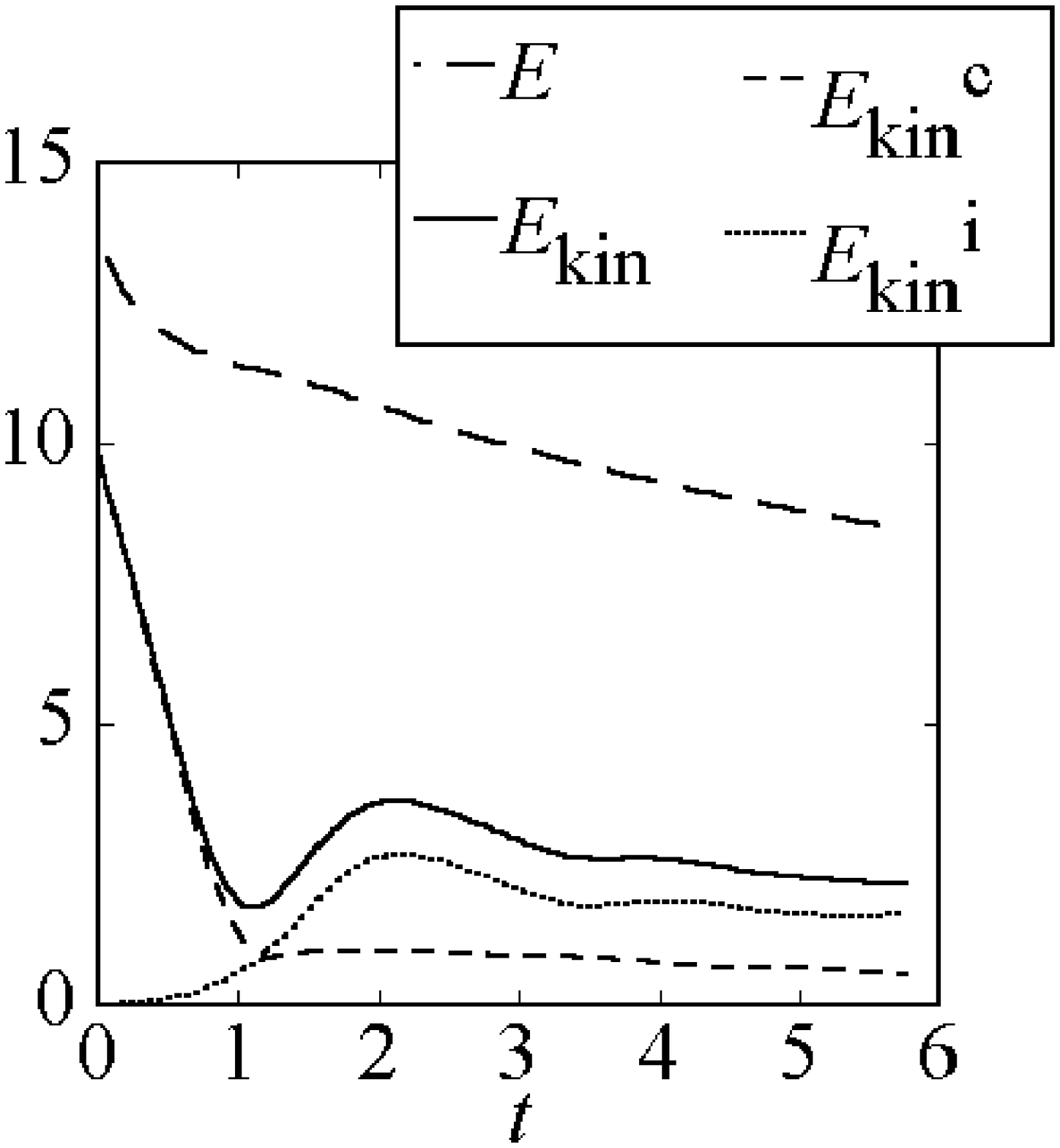}\\
(b)
\end{center}
\end{minipage}
\caption{\label{fig-gammaE} Time development of $E$, $E\sub{kin}$, $E\sub{kin}\up{c}$ and $E\sub{kin}\up{i}$.
(a) Case with $\gamma_0=0$ (b) case with $\gamma_0=1$.}
\end{center}
\end{figure}

We introduced a dissipation term in Eq. (\ref{eq-FtransGP}) to remove excitations of wavelength shorter than the
 healing length.
The imaginary number $\ii$ in the left-hand side of Eq. (\ref{eq-FtransGP}) was replaced by $[\ii-\gamma(k)]$, where
 $\gamma(k)=\gamma_0\theta(k-2\pi/\xi)$ with $\theta(x)$ being the step function.
The effect of this dissipation is shown in Fig. \ref{fig-gammaE}.
We divide up to the kinetic energy $E\sub{kin}$ into the compressible part
 $E\sub{kin}\up{c}=\int\dd\Vec{x}\:[(|\Phi|\nabla\phi)\up{c}]^2/\int\dd\Vec{x}\:\rho$, due to sound waves, and the
 incompressible part $E\sub{kin}\up{i}=\int\dd\Vec{x}\:[(|\Phi|\nabla\phi)\up{i}]^2/\int\dd\Vec{x}\:\rho$, due to
 vortices, where $\rot (|\Phi|\nabla\phi)\up{c}=0$ and $\diver (|\Phi|\nabla\phi)\up{i}=0$ \cite{Nore,Ogawa}.
Figure \ref{fig-gammaE} shows the time development of $E$, $E\sub{kin}$, $E\sub{kin}\up{c}$ and $E\sub{kin}\up{i}$ in
 the case of $\gamma_0=0$ (a) and $\gamma_0=1$ (b).
Without dissipation, the compressible kinetic energy $E\sub{kin}\up{c}$ is increased in spite of conserving the total
 energy  (Fig. \ref{fig-gammaE}(a)), which is consistent with the simulation by Nore {\it et al}. \cite{Nore}.
The dissipation suppresses $E\sub{kin}\up{c}$ and thus causes $E\sub{kin}\up{i}$ to be dominant.
This dissipation term works as viscosity only at the small scale.
At other scales, which are equivalent to the inertial range $\Delta k<k<2\pi/\xi$, the energy does not dissipate.

\begin{figure}[htb]
\begin{center}
\begin{minipage}{0.49\linewidth}
\begin{center}
\includegraphics[width=0.95\linewidth]{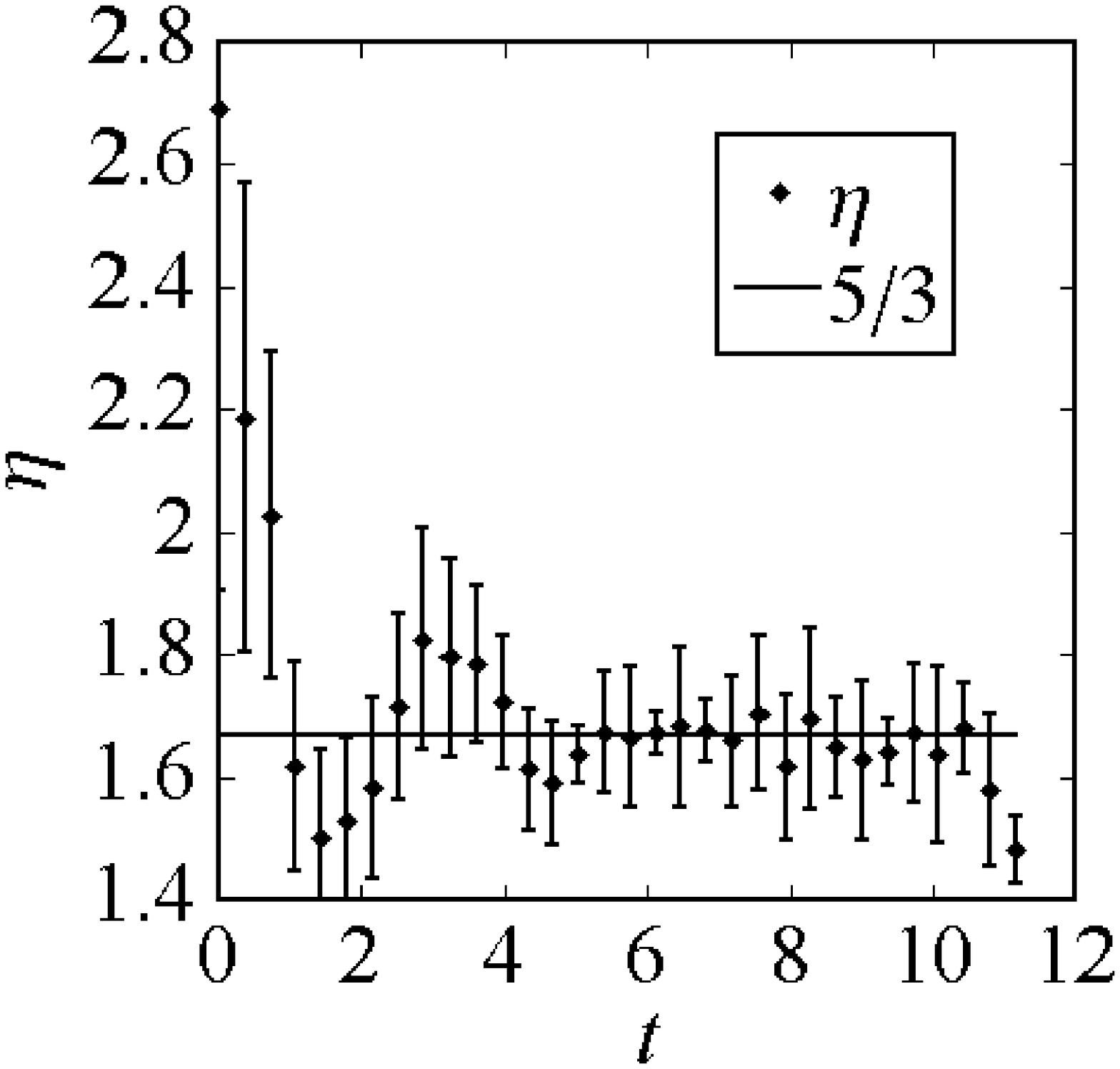}\\
(a)
\end{center}
\end{minipage}
\begin{minipage}{0.49\linewidth}
\begin{center}
\includegraphics[width=0.95\linewidth]{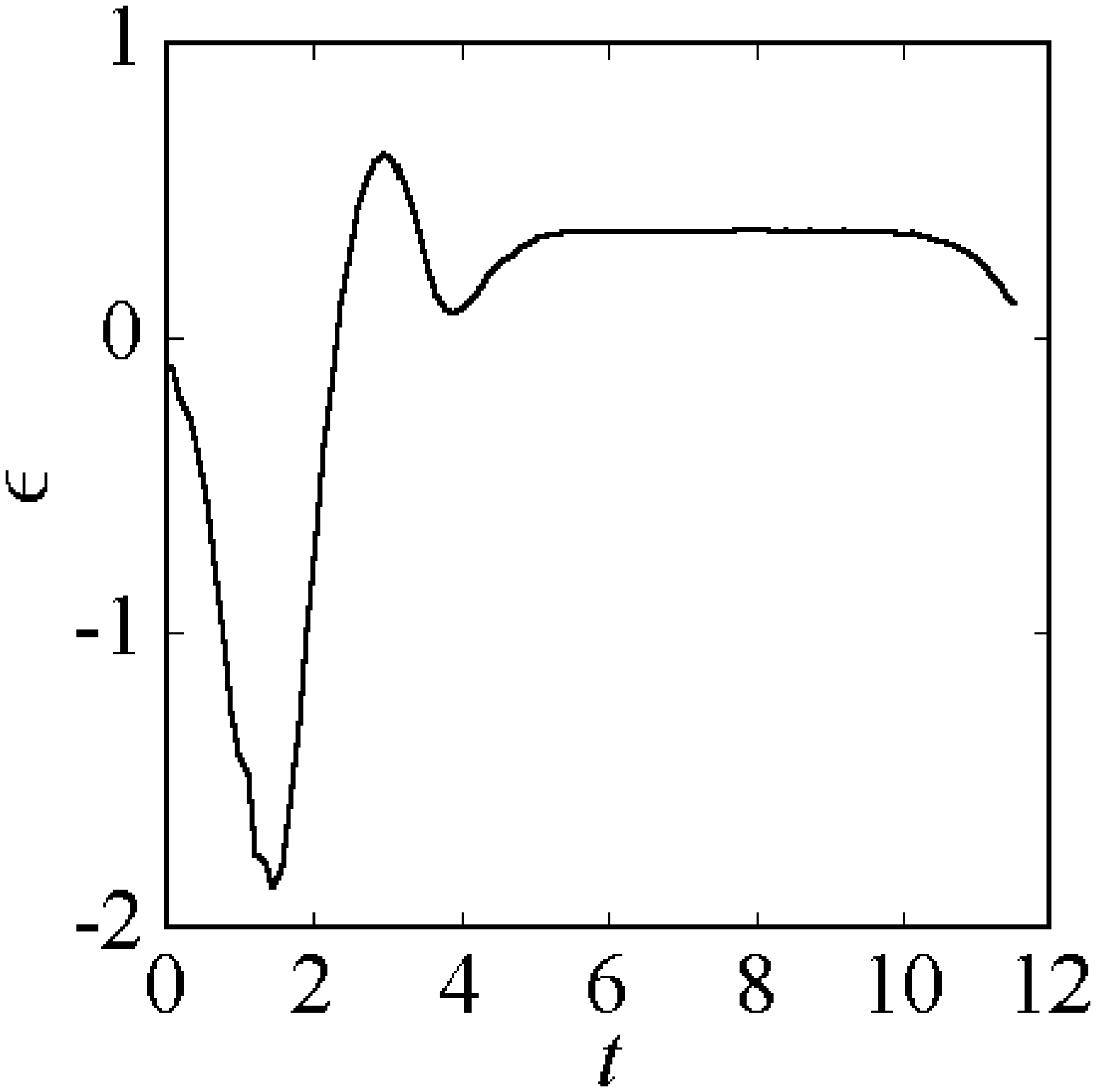}\\
(b)
\end{center}
\end{minipage}\\
\begin{minipage}{0.49\linewidth}
\begin{center}
\includegraphics[width=0.95\linewidth]{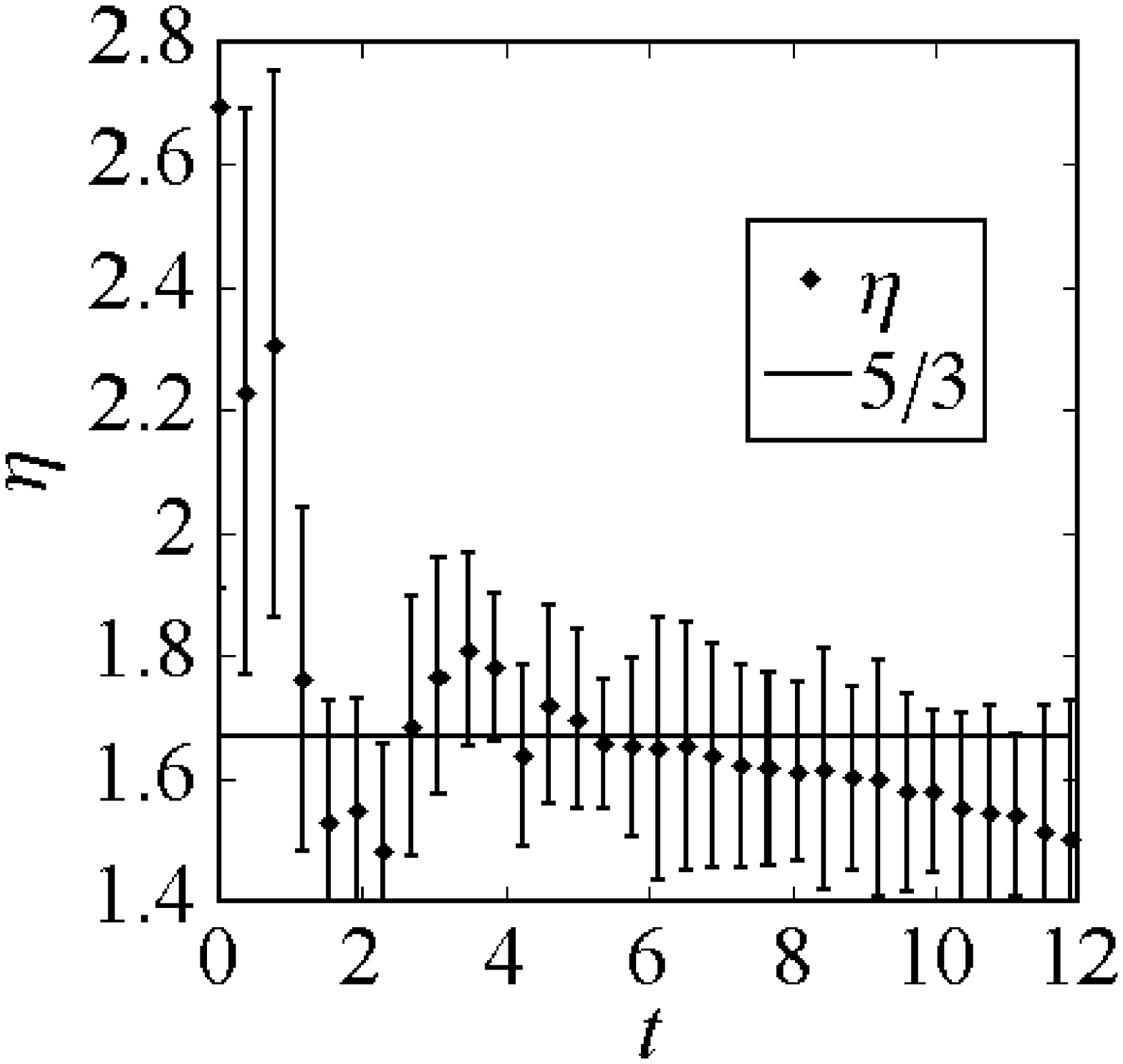}\\
(c)
\end{center}
\end{minipage}
\begin{minipage}{0.49\linewidth}
\begin{center}
\includegraphics[width=0.95\linewidth]{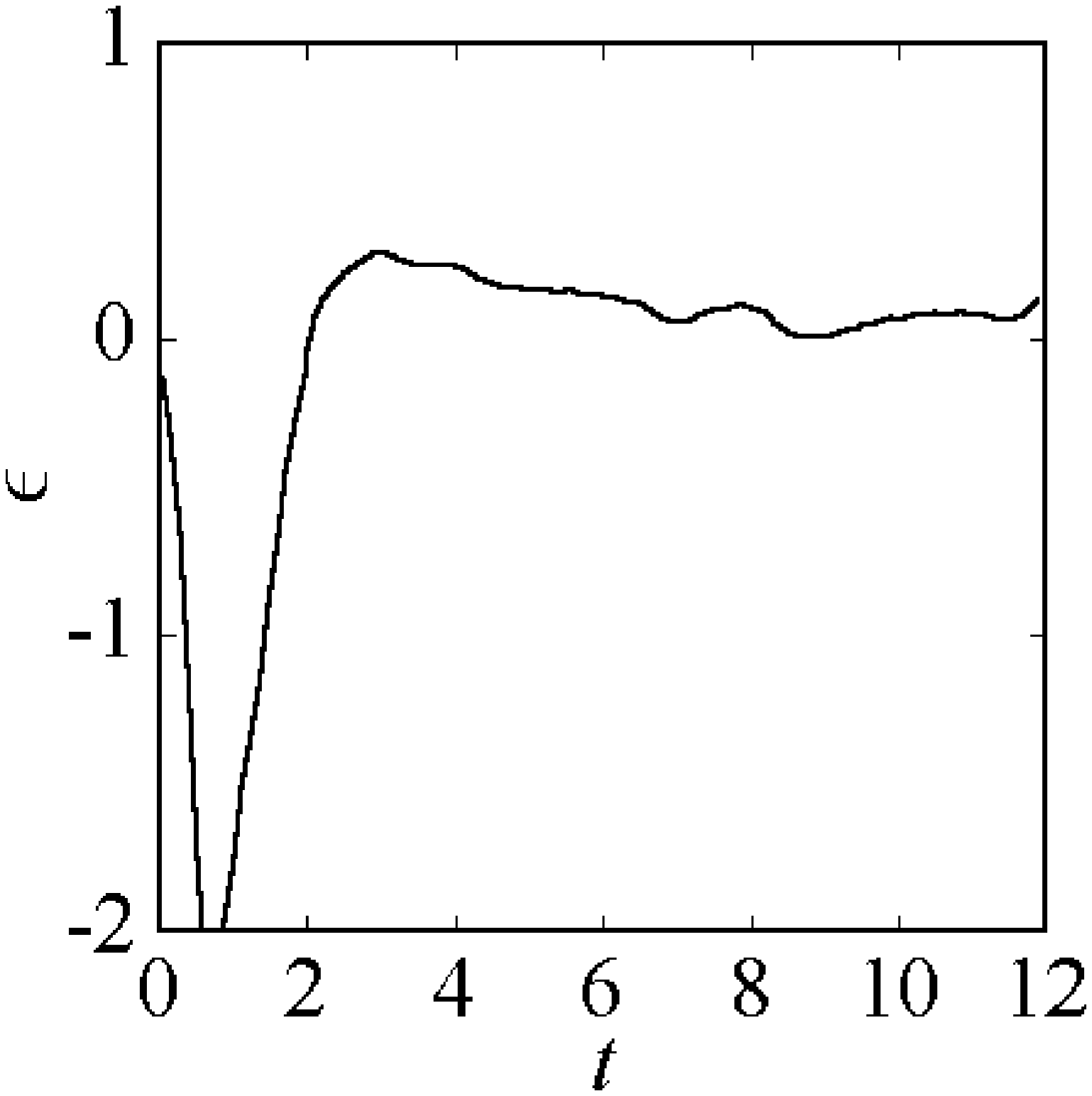}\\
(d)
\end{center}
\end{minipage}
\caption{\label{fig-exp} Time dependence of the exponent of the spectrum $\eta$ (a) and the dissipation rate
 $\epsilon$.
(a) $\eta$ for $\gamma_0=1$.
(b) $\epsilon$ for $\gamma_0=1$.
(c) $\eta$ for $\gamma_0=0$.
(d) $\epsilon$ for $\gamma_0=0$.
In (a) and (c), the error bars are the standard deviation of the data from the fit.
The line $\eta=5/3$ is shown to compare the results with the Kolmogorov law.}
\end{center}
\end{figure}

We calculated the spectrum of the incompressible kinetic energy $E\sub{kin}\up{i}(k)$ defined as
 $E\sub{kin}\up{i}=\int\dd k\: E\sub{kin}\up{i}(k)$.
Initially, the spectrum $E\sub{kin}\up{i}(k)$ significantly deviates from the Kolmogorov power-law, however, the
 spectrum approaches a power-law as the turbulence develops.
We assumed that the spectrum $E\sub{kin}\up{i}(k)$ is proportional to $k^{-\eta}$ in the inertial range
 $\Delta k<k<2\pi/\xi$, and we determined the exponent $\eta$.
The time development of $\eta$ is shown in Fig. \ref{fig-exp} (a).
This figure shows that our ST satisfies the Kolmogorov law $E\sub{kin}\up{i}(k)\propto k^{-5/3}$
 for times $4\lesssim t\lesssim 10$, when the system may be almost homogeneous and isotropic turbulence.
We also calculated the energy dissipation rate $\epsilon=-\partial E\sub{kin}\up{i}/\partial t$ and compared the
 results quantitatively with the Kolmogorov law.
Figure \ref{fig-exp} (b) shows that $\epsilon$ is almost constant in the period $4\lesssim t\lesssim 10$,
 which means that the Kolmogorov spectrum $\epsilon^{2/3}k^{-\eta}$ is also constant in the period.
Without the dissipation term $\gamma(k)$, the ST could not satisfy the Kolmogorov law.
This means that dissipating short wavelength excitations are essential for satisfying the Kolmogorov law.
Figure \ref{fig-exp} (c) shows the time development of the exponent $\eta$ for $\gamma_0=0$.
In this case, $\eta$ is consistent with the Kolmogorov law only in the short period $4\lesssim t\lesssim 7$; active
 sound waves affect the dynamics of quantized vortices for $t\gtrsim 7$.
Such an effect of sound waves is also clearly shown in Fig. \ref{fig-exp} (d).
Compared with Fig. \ref{fig-exp} (b), $\epsilon$ is unsteady and smaller than that of $\gamma_0=1$ and even becomes
 negative at times.
When $\epsilon$ is negative, the energy flows backward from sound waves to vortices, which may prevent the energy
 spectrum from satisfying the Kolmogorov law for $t\gtrsim 7$.
According to Fig. \ref{fig-exp} (a), the agreement between the energy spectrum and the Kolmogorov law becomes weak at
 a later stage $t\gtrsim 10$, which may be attributable to the following reasons.
In the period $4\lesssim t\lesssim 10$, the energy spectrum agrees with the Kolmogorov law, which may support that the
 Richardson cascade process works in the system.
The dissipation is caused mainly by removing short wavelength excitations emitted at vortex reconnections.
However, the system at the late stage $t\gtrsim 10$ has only small vortices after the Richardson cascade process, being
 no longer turbulent.
The energy spectrum, therefore, disagrees with the Kolmogorov law of $\eta=5/3$.
And emissions of excitations through vortex reconnections hardly occurs, which reduces greatly the energy dissipation
 rate $\epsilon$.

\begin{figure}[htb]
\begin{center}
\begin{minipage}{0.49\linewidth}
\begin{center}
\includegraphics[width=0.95\linewidth]{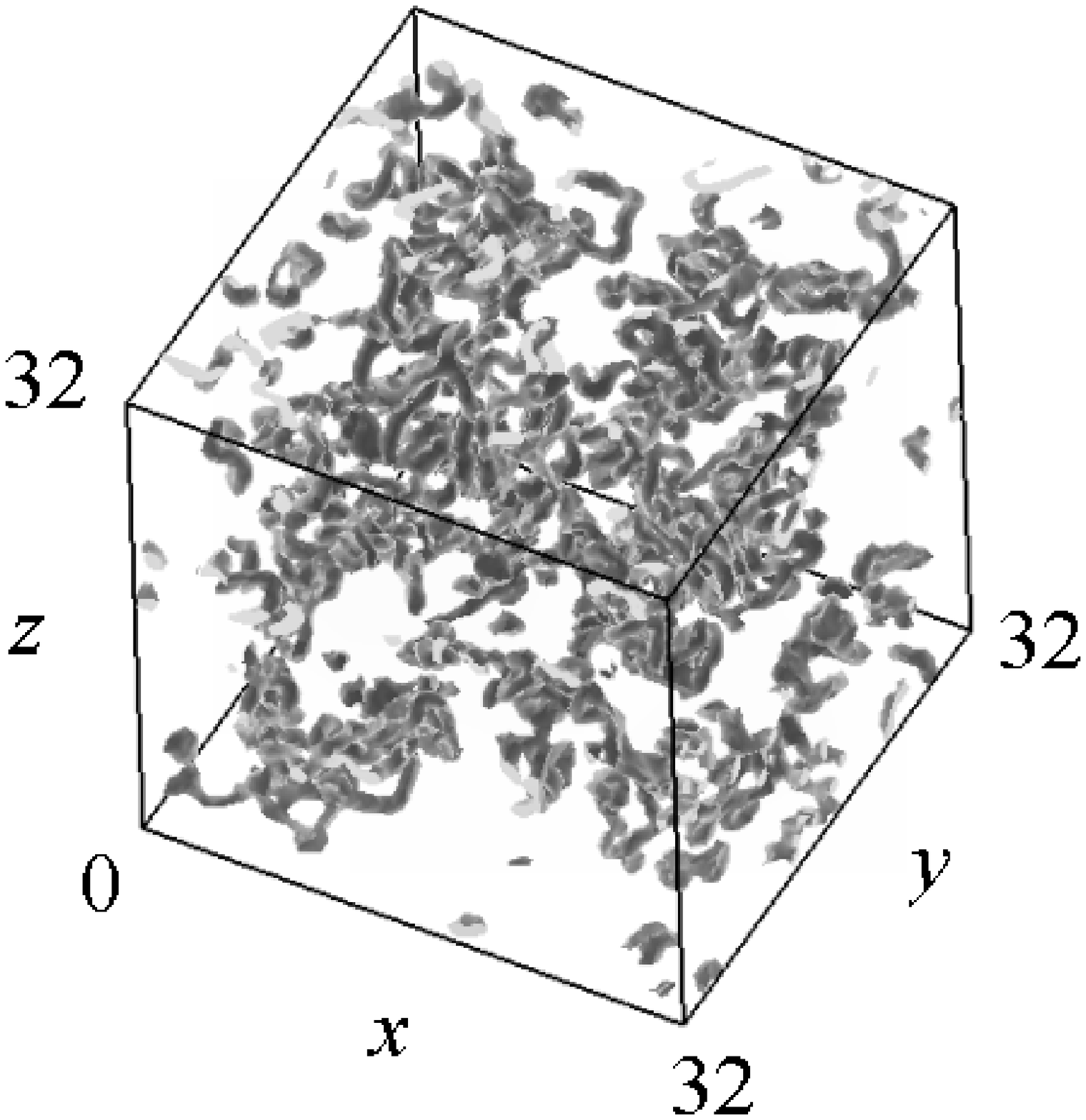}\\
(a)
\end{center}
\end{minipage}
\begin{minipage}{0.49\linewidth}
\begin{center}
\includegraphics[width=0.95\linewidth]{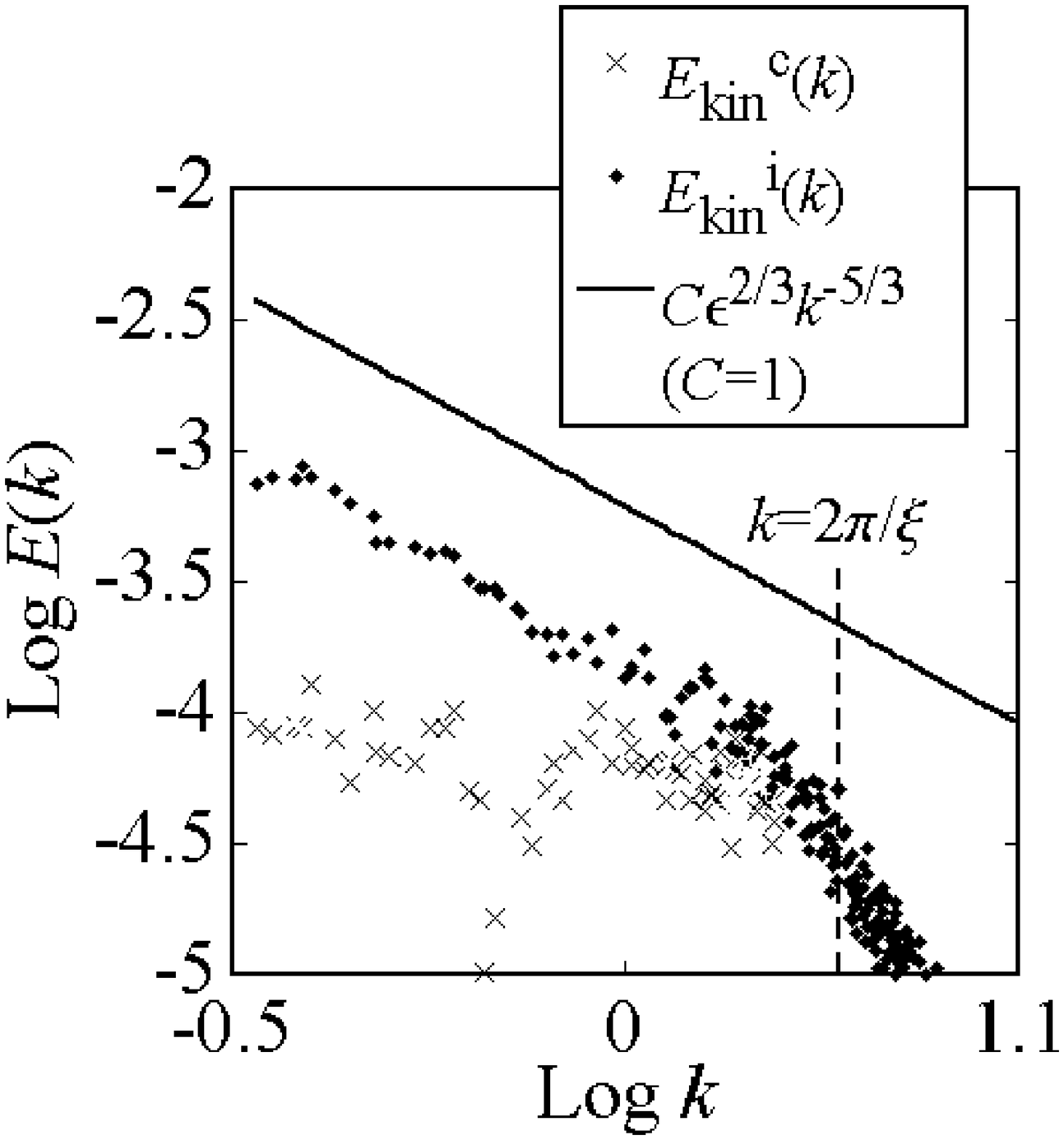}\\
(b)
\end{center}
\end{minipage}
\caption{\label{fig-Kolmo} Vorticity $\rot\Vec{v}(\Vec{x},t)$ and the energy spectrum at $t=6.0$.
The contour in (a) is 98\% of maximum vorticity.
The energy spectrum was obtained by making an ensemble average for 20 different initial states
The solid line refers to the Kolmogorov law in Eq. (\ref{eq-Kolmo}).}
\end{center}
\end{figure}

Spatially, the system appears to develop full turbulence.
Shown in Fig. \ref{fig-Kolmo} (a) is the spatial distribution of vortices.
The vorticity plot suggests fully developed turbulence.
The energy spectrum in Fig. \ref{fig-Kolmo} (b) agrees quantitatively with the Kolmogorov law.
We thus conclude that the incompressible kinetic energy of ST without the effect of short wavelength excitations
 satisfies the Kolmogorov law.
This inertial range $0.20\lesssim k\lesssim 6.3$ is larger than those in the simulations by Araki {\it et al}.
 \cite{Araki} and Nore {\it et al}. \cite{Nore}.
The Kolmogorov constant is estimated to be $C\simeq 10^{-0.5}\sim 0.32$.
This Kolmogorov constant is small because this turbulence is dissipative, being expected to become larger for a steady
 turbulence with injection of energy.

In conclusion, we investigate the energy spectrum of ST by the numerical simulation of the GP equation.
Suppressing the compressible sound waves which are created through vortex reconnections, we can make the quantitative
 agreement between the spectrum of incompressible kinetic energy and Kolmogorov law.
Our next subjects are making the steady turbulence by introducing injection of the energy and broadening the
 inertial range of the energy spectrum.
We will report on this results in near future.

We acknowledge W. F. Vinen for useful discussions.
MT and MK both acknowledge support by a Grant-in-Aid for Scientific Research (Grant No.15340122 and 1505983
 respectively) by the Japan Society for the Promotion of Science. 

\end{document}